\documentclass[12pt]{article}
\usepackage{graphicx}
\usepackage{subfig}
\usepackage{epsfig}
\usepackage{rotating}
\usepackage{amsmath}
\def\br(#1,#2){\left\langle#1#2\right\rangle}
\def\sq(#1,#2){\left[#1#2\right]}
\def\s(#1,#2){s_{#1 #2}}
\def\t(#1,#2,#3){s_{#1 #2 #3}}

\begin{document}

\begin{titlepage}

\hspace*{\fill}\parbox[t]{5cm}
{\today \\
MPP-2011-149 \\
SI-HEP-2011-17\\
CP3-12-25} \vskip2cm
\begin{center}
{\Large \bf Effective Field Theory: \\
\bigskip  A Modern Approach to Anomalous Couplings} \\
\medskip
\bigskip\bigskip\bigskip\bigskip
{\bf C\'eline Degrande,$^{1,2}$ Nicolas Greiner,$^{1,3}$ Wolfgang Kilian,$^{1,4}$ Olivier Mattelaer,$^2$ \\
\medskip
Harrison Mebane,$^1$ Tim Stelzer,$^1$  Scott Willenbrock,$^1$ and Cen Zhang$^{1,2}$}\\
\bigskip\bigskip\bigskip
$^1$Department of Physics, University of Illinois at Urbana-Champaign \\ 1110 West Green Street, Urbana, IL  61801 \\
\bigskip
$^2$Institut de Physique Th\'eorique and Centre for Particle Physics and Phenomenology (CP3) \\ Universit\'e Catholique de Louvain \\ Chemin du Cyclotron 2, B-1348 Louvain-la-Neuve, Belgium \\
\bigskip
$^3$Max-Planck-Institut f\"ur Physik, F\"ohringer Ring 6, 80805 M\"unchen, Germany \\
\bigskip
$^4$University of Siegen, Fachbereich Physik, D-57068 Siegen, Germany
\end{center}

\bigskip\bigskip\bigskip

\begin{abstract}
We advocate an effective field theory approach to anomalous couplings. The effective field theory approach is the natural way to extend the standard model such that the gauge symmetries are respected. It is general enough to capture any physics beyond the standard model, yet also provides guidance as to the most likely place to see the effects of new physics.  The effective field theory approach also clarifies that one need not be concerned with the violation of unitarity in scattering processes at high energy.  We apply these ideas to pair production of electroweak vector bosons.
\end{abstract}

\end{titlepage}

\section{Introduction}\label{sec:intro}

In this paper we compare and contrast two conceptual frameworks that were contemporaneously introduced into particle physics in the late 1970's: effective field theory \cite{Weinberg:1978kz,Weinberg:1979pi} and anomalous couplings of electroweak vector bosons \cite{Gaemers:1978hg}.  We argue in favor of the effective-field-theory approach, and suggest that the time has come to retire the anomalous-coupling framework.  Although we concentrate on the case of electroweak vector bosons, our arguments apply to all anomalous-coupling paradigms.

The anomalous couplings of electroweak vector bosons were introduced at a time when it was not certain that the electroweak interaction was a spontaneously broken gauge theory (although it was widely believed to be the case) \cite{Gaemers:1978hg}.  Thus there was some motivation to extend the framework of the standard electroweak theory \cite{Weinberg:1967tq,Salam} to allow for vector-boson interactions that differed from the gauge-theory prescription \cite{Gaemers:1978hg,Hagiwara:1986vm}.  That motivation has evaporated now that we have a wide variety of precision measurements of electroweak processes.  It is overly permissive to consider an extension of the standard electroweak model that does not incorporate $SU(3)_C\times SU(2)_L\times U(1)_Y$ gauge symmetry.  In particular, there is no well-defined prescription to employ anomalous couplings in the loop calculations that are necessary to describe the precision electroweak data.  In contrast, the effective-field-theory framework allows for the unambiguous calculation of loop effects \cite{De Rujula:1991se,Hagiwara:1992eh,Hagiwara:1993ck}.  While anomalous couplings were perhaps useful at one time, they are now much too crude to be used to confront theory with experiment.

One of the unpleasant complications of anomalous couplings, even at tree level, is that they are often promoted from constants to arbitrarily-chosen form factors \cite{Zeppenfeld:1987ip,Baur:1987mt,Hagiwara:1989mx}.  We argue that this complication is absent in the effective-field-theory approach.  Thus the effective-field-theory framework is also cleaner and simpler than that of anomalous couplings.

Effective field theory ideas have already been used in the context of anomalous couplings \cite{De Rujula:1991se,Hagiwara:1992eh,Hagiwara:1993ck,Wudka:1994ny,Willenbrock:1995kf,Ellison:1998uy}, but these ideas have not been fully embraced by the high-energy community.
We hope to convince the reader that the time is ripe to adopt the effective field theory approach and abandon the anomalous couplings framework.

\section{Effective Field Theory}\label{sec:eft}

There are two methods to search for physics beyond the standard model.  One is to look for the new physics directly, via the production of new particles.  The other is to look for novel interactions of the known particles of the standard model.  Here we are focussed on the latter method.

We desire to take a model-independent approach to the physics of nonstandard interactions.
A model-independent approach is useful in two respects.  First, it allows one to search for new physics without committing to a particular extension of the standard model.  Second, in the case that no new physics should appear, it allows one to quantify the accuracy with which the new physics is excluded.

When contemplating a model-independent approach to nonstandard interactions, there are a number of desirable features that one should incorporate:
\begin{itemize}
\item Any extension of the standard model should satisfy the $S$-matrix axioms of unitary, analyticity, {\it etc.}
\item The symmetries of the standard model, namely Lorentz invariance and $SU(3)_C\times SU(2)_L\times U(1)_Y$ gauge symmetry, should be respected.
\item It should be possible to recover the standard model in an appropriate limit.
\item The extended theory should be general enough to capture any physics beyond the standard model, but should give some guidance as to the most likely place to see the effects of new physics.
\item It should be possible to calculate radiative corrections at any order in the standard-model interactions in the extended theory.
\item It should be possible to calculate radiative corrections at any order in the new interactions of the extended theory.
\end{itemize}
The unique way to incorporate all of these features is via an effective quantum field theory. The first two features alone indicate a quantum field theory.  The remaining features are captured by an effective quantum field theory \cite{Weinberg:1978kz,Weinberg:1979pi}.

An effective quantum field theory of the standard model is constructed as follows.  The standard model is the most general theory of quark and lepton fields, along with a single Higgs doublet field, interacting via an $SU(3)_C\times SU(2)_L\times U(1)_Y$ gauge symmetry, where all operators (that is, products of fields) in the Lagrangian are restricted to be of mass dimension four or less.\footnote{In practice all operators, except the quadratic term in the Higgs potential, are of dimension four.}  To extend the theory, add operators of higher dimension.  By dimensional analysis, these operators have coefficients of inverse powers of mass, and hence are suppressed if this mass is large compared with the experimentally-accessible energies.  The dominant extended operators will therefore be those of the lowest dimensionality.  There is only one operator of mass dimension five, and it is responsible for generating Majorana masses for neutrinos \cite{Weinberg:1979sa}.  There are many operators of dimension six, but typically only a few contribute to a given physical process \cite{Leung:1984ni,Buchmuller:1985jz}.  It is the dimension-six operators that will concern us throughout this article.

As is customary, we will denote the mass scale that characterizes the coefficients of the higher-dimension operators as $\Lambda$. It can be regarded as the scale of new physics.  The underlying assumption of an effective quantum field theory is that this scale is large compared with the experimentally-accessible energies.  Thus an effective field theory is the low-energy approximation to the new physics, where ``low'' means less than $\Lambda$.  The scale $\Lambda$ could be anywhere from about a TeV up to the Planck scale.  The framework is so general that the new physics could be anything, including new particles, extra spacetime dimensions, or even physics that is not described by ordinary quantum field theory (such as string theory).

The effective quantum field theory of the standard model is (ignoring the one dimension-five operator involving only lepton and Higgs fields)
\begin{equation}
{\cal L} = {\cal L}_{SM} + \sum_i \frac{c_i}{\Lambda^2}{\cal O}_i+\cdots
\label{eq:eft}
\end{equation}
where ${\cal O}_i$ are the dimension-six operators, and the ellipsis indicates the yet higher-dimension operators.\footnote{Only even-dimensional operators conserve both lepton and baryon number.  A proof is provided in Appendix~\ref{apdx:dimodd}.} The coefficients $c_i$ are dimensionless, and parameterize the strength with which the new physics couples to the standard-model particles.

An example of an effective field theory is provided by extending the standard model to include a heavy $Z^\prime$ boson.  At energies less than the $Z^\prime$ mass, one will not observe the $Z^\prime$ directly, but one can see its indirect effects.  For example, the exchange of a $Z^\prime$ boson between two standard-model fermion currents would appear as a four-fermion interaction at low energies, that is, at energies less than the $Z^\prime$ mass.  This is a dimension-six operator.  It will be suppressed by two inverse powers of the $Z^\prime$ mass, coming from the $Z^\prime$ propagator; thus $\Lambda = M_{Z^\prime}$.  The dimensionless coefficient $c$ will be the product of the couplings of the standard-model fermions to the $Z^\prime$.  This example is analogous to the way the Fermi theory of the weak interaction arises from the Standard Model at energies below the $W$ boson mass.

Referring to our list of desirable features above, we see that the standard model is recovered in the limit $\Lambda \to \infty$.  Since any new physics will look like a quantum field theory at low energies, the effective field theory is general enough to capture the low-energy effects of any physics beyond the standard model, as long as we include all possible terms consistent
with the symmetries of the theory.  However, by dimensional analysis we expect the dimension-six operators to be dominant, so the theory provides some guidance as to the most likely place to see the effects of new physics.\footnote{For some physical processes, operators of dimension seven or greater may be dominant, and can be included.}  Finally, the extended theory can be used to calculate both tree-level and loop processes \cite{Gomis:1995jp}.

In the effective field theory approach one must completely specify the particle content of the theory, that is, the fields from which the operators ${\cal O}_i$ are constructed.  These are the fields present at low energies, that is, energies less than $\Lambda$.  We assume that these are the fields of the standard model, including the usual Higgs doublet field, and nothing more.  We include the Higgs doublet field because precision electroweak data support the presence of a Higgs boson at low energy \cite{Nakamura:2010zzi}.\footnote{If the physics of electroweak symmetry breaking is at a scale much greater than the $W$ mass, then an effective field theory with $SU(2)_L\times U(1)_Y$ realized nonlinearly is indicated \cite{Appelquist:1980vg,Longhitano:1980iz,Longhitano:1980tm}.  The application of such an approach to electroweak vector boson pair production is pursued in Ref.~\cite{Falk:1991cm}.}  If future experiments should reveal new particles at low energies (including an extended Higgs sector), then one would need to revise the effective field theory to include the associated fields.


An effective quantum field theory is only useful up to energies of order $\Lambda$.  At energies beyond that, there is no justification for neglecting operators of dimension greater than six, since they are not suppressed.  Operators of arbitrarily high dimension become important, and since there are an infinite number of them the usefulness of the approach is lost.  Our discussion of a $Z^\prime$ boson above provides an example.  As the experimentally-accessible energies approach the $Z^\prime$ mass, the yet higher-dimension operators are no longer suppressed, and the effective field theory becomes useless.  It is replaced by a new theory which includes a $Z^\prime$ among its low-energy content.  One can then develop a new effective field theory, including the $Z^\prime$, to parameterized the low-energy effects of new physics at a yet higher scale.  Thus the effective field theory approach is adaptable as we discover new heavy particles.

\section{Dimension-Six Operators}\label{sec:dimsix}

There are many dimension-six operators, but only a few of them affect any given physical process \cite{Buchmuller:1985jz}.  Thus, by making a variety of measurements, one can measure or constrain many of the coefficients of these operators.  The number of independent $B$- and $L$-conserving dimension-six operators is 59 for one generation of quarks and leptons \cite{Grzadkowski:2010es}.

Electroweak vector boson pair production involves the coupling of the electroweak vector bosons to fermions and to each other.  The coupling of the electroweak vector bosons to fermions is generally constrained by other processes, so it is reasonable to focus on the electroweak vector boson self interactions when considering the contribution of dimension-six operators to electroweak vector boson pair production. Assuming $C$ and $P$ conservation, there are just three independent dimension-six operators that affect the electroweak vector boson self interactions \cite{Hagiwara:1993ck}.  There is some flexibility in which three operators are chosen.  We follow Ref.~\cite{Hagiwara:1993ck} and chose the three independent $C$ and $P$ conserving operators to be
\begin{eqnarray}
{\cal O}_{WWW}&=&\mbox{Tr}[W_{\mu\nu}W^{\nu\rho}W_{\rho}^{\mu}]\\
{\cal O}_W&=&(D_\mu\Phi)^\dagger W^{\mu\nu}(D_\nu\Phi)\label{eq:OW}\\
{\cal O}_B&=&(D_\mu\Phi)^\dagger B^{\mu\nu}(D_\nu\Phi)
\end{eqnarray}
where $\Phi$ is the Higgs doublet field and
\begin{align}
D_\mu & = \partial_\mu + \frac{i}{2} g \tau^I W^I_\mu + \frac{i}{2} g' B_\mu \\
W_{\mu\nu} & = \frac{i}{2} g\tau^I (\partial_\mu W^I_\nu - \partial_\nu W^I_\mu
	+ g \epsilon_{IJK} W^J_\mu W^K_\nu )\\
B_{\mu \nu} & = \frac{i}{2} g' (\partial_\mu B_\nu - \partial_\nu B_\mu)
\end{align}
This is a good choice of operators as they are constrained only by electroweak vector boson pair production
\cite{Grojean:2006nn}.

There is no reason to believe that $C$ and $P$ are conserved by the dimension-six operators, unless the physics beyond the standard model respects these symmetries.  If we allow for $C$ and/or $P$ violation, there are two additional operators, which we choose to be
\begin{eqnarray}
{\cal O}_{\tilde WWW}&=&\mbox{Tr}[{\tilde W}_{\mu\nu}W^{\nu\rho}W_{\rho}^{\mu}]\\
{\cal O}_{\tilde W}&=&(D_\mu\Phi)^\dagger {\tilde W}^{\mu\nu}(D_\nu\Phi)
\end{eqnarray}
Thus there are three $C$ and $P$ conserving dimension-six operators and two operators that violate $C$ and/or $P$.  Together these five operators parameterize the leading effect of physics beyond the standard model on the electroweak vector boson self interactions.

\section{Anomalous Couplings}\label{sec:anomalous}

Anomalous couplings of electroweak vector bosons are discussed in one of two formalisms: a Lagrangian or a vertex function.  Here we discuss these two approaches and compare them with the effective field theory approach discussed in the previous sections.

\subsection{Lagrangian approach}

The Lagrangian approach to anomalous couplings is based on the Lagrangian \cite{Hagiwara:1986vm}
\begin{eqnarray}
{\cal L}&=&ig_{WWV}\left(g_1^V(W_{\mu\nu}^+W^{-\mu}-W^{+\mu}W_{\mu\nu}^-)V^\nu
+\kappa_VW_\mu^+W_\nu^-V^{\mu\nu}
+\frac{\lambda_V}{M_W^2}W_\mu^{\nu+}W_\nu^{-\rho}V_\rho^{\mu}
\right.\nonumber\\&&\left.
+ig_4^VW_\mu^+W^-_\nu(\partial^\mu V^\nu+\partial^\nu V^\mu)
-ig_5^V\epsilon^{\mu\nu\rho\sigma}(W_\mu^+\partial_\rho W^-_\nu-\partial_\rho W_\mu^+W^-_\nu)V_\sigma
\right.\nonumber\\&&\left.
+\tilde{\kappa}_VW_\mu^+W_\nu^-\tilde{V}^{\mu\nu}
+\frac{\tilde{\lambda}_V}{m_W^2}W_\mu^{\nu+}W_\nu^{-\rho}\tilde{V}_\rho^{\mu}
\right)
\label{eq:L}
\end{eqnarray}
where $V=\gamma,Z$;  $W_{\mu\nu}^\pm = \partial_\mu W_\nu^\pm - \partial_\nu W_\mu^\pm$, $V_{\mu\nu} = \partial_\mu V_\nu - \partial_\nu V_\mu$, and the overall coupling constants are defined as $g_{WW\gamma}=-e$ and $g_{WWZ}=-e\cot\theta_W$.  In constructing this Lagrangian, the $W$ bosons are constrained to be on shell and the scalar components of the neutral gauge bosons are neglected \cite{Hagiwara:1986vm}.  These are appropriate constraints for application of this Lagrangian to electroweak vector boson pair production, so they do not imply a loss of generality.  None of these constraints need be imposed on the effective field theory, however; it is valid for both real and virtual particles, and for all field components.

The first three terms of Eq.~(\ref{eq:L}) respect $C$ and $P$, and the remaining four terms violate $C$ and/or $P$.
Electromagnetic gauge invariance implies that $g_1^\gamma =1$ and $g_4^\gamma=g_5^\gamma = 0$.  Thus there are five independent $C$- and $P$-conserving parameters: $g_1^Z, \kappa_\gamma, \kappa_Z, \lambda_\gamma, \lambda_Z$; and six $C$ and/or $P$ violating parameters: $g_4^Z, g_5^Z, \tilde{\kappa}_\gamma, \tilde{\kappa_Z}, \tilde{\lambda}_\gamma, \tilde{\lambda_Z}$

This is the most general Lagrangian describing the trilinear interaction of electroweak vector bosons, but only in a limited sense.  The Lagrangian contains all possible Lorentz structures, each constructed with the fewest number of derivatives.  However, one can construct an infinite number of additional terms by adding derivatives, $\partial_\mu$  \cite{Hagiwara:1986vm}.  Each derivative would be accompanied by a factor of $M_W^{-1}$, since this is the only mass in the theory (cf.\ the two dimension six terms proportional to $\lambda_V$ and $\tilde\lambda_V$).  These additional terms are not suppressed at energies above the $W$ mass, and thus there is no principle instructing one to neglect them.
This is in contrast to the effective field theory approach, where each additional power of $D_\mu$ is accompanied by a factor of $\Lambda^{-1}$, and is hence suppressed.  If one were to argue that only the lowest dimension operators should be kept (as in an effective field theory), then there would be no rationale for keeping the two dimension six terms proportional to $\lambda_V$ and $\tilde\lambda_V$ and dropping terms constructed from the other terms with two derivatives added, which are also dimension six.  This Lagrangian lacks the systematic expansion in powers of $\Lambda^{-1}$ present in an effective field theory.

This is not the Lagrangian of a modern effective field theory. No attention has been paid to the issue of $SU(2)_L\times U(1)_Y$ gauge symmetry, despite the fact that the Lagrangian is constructed from the electroweak vector bosons.  If this Lagrangian is used for tree-level calculations, it leads to violation of unitarity bounds at high energy, with no prescription for how to deal with them \cite{Zeppenfeld:1987ip,Baur:1987mt,Hagiwara:1989mx}.  The issue of unitarity bounds is discussed in Section \ref{sec:unitarity}.  In loop calculations, it generally yields ultraviolet divergences, again with no prescription for how to handle them
\cite{De Rujula:1991se,Hagiwara:1992eh,Hagiwara:1993ck}.  Thus this Lagrangian lacks many of the virtues of an effective field theory.

\subsection{Vertex function approach}

The vertex function approach is the momentum-space analogue of the Lagrangian approach.
The vertex-function approach parameterizes the trilinear vector boson vertex function as \cite{Gaemers:1978hg,Hagiwara:1986vm}
\begin{eqnarray}
\Gamma_V^{\alpha\beta\mu}&=&f_1^V(q-\bar q)^\mu g^{\alpha\beta}-\frac{f_2^V}{M_W^2}(q-\bar q)^\mu P^\alpha P^\beta
+f_3^V(P^\alpha g^{\mu\beta}-P^\beta g^{\mu\alpha})
\nonumber\\&&
+if_4^V(P^\alpha g^{\mu\beta}+P^\beta g^{\mu\alpha})
+if_5^V\epsilon^{\mu\alpha\beta\rho}(q-\bar{q})_\rho
\nonumber\\&&
-f_6^V\epsilon^{\mu\alpha\beta\rho}P_\rho
-\frac{f_7^V}{m_W^2}(q-\bar{q})^\mu
\epsilon^{\alpha\beta\rho\sigma}P_\rho (q-\bar{q})_\sigma
\end{eqnarray}
where $P,q,\bar q$ are the four-momenta of $V,W^-,W^+$, respectively.  The $W$ bosons are constrained to be on shell, and the scalar components of the neutral gauge bosons are neglected, but as in the Lagrangian approach this does not imply a loss of generality when applied to electroweak vector boson pair production.  The first three terms respect $C$ and $P$, and the remaining four terms violate $C$ and/or $P$.

The coefficients $f_i^V$ are form factors that depend on $P^2$.  The functional form of the $f_i^V(P^2)$ is arbitrary.  This is the momentum space analogue of the infinite number of terms in the Lagrangian approach that can be constructed by including more derivatives.
The $W$ boson electric charge implies that $f_1^\gamma(0)=1$, and the electromagnetic Ward identity implies that $f_4^\gamma, f_5^\gamma$ are proportional to $P^2$ \cite{Hagiwara:1986vm}.  These are the momentum space analogues of the constraints imposed by electromagnetic gauge invariance in the Lagrangian approach.

As in the Lagrangian approach to anomalous couplings, no attention has been paid to the issue of $SU(2)_L\times U(1)_Y$ gauge invariance.  The form factors are often chosen to be suppressed at large $P^2$ in order to avoid violation of unitarity bounds, but there is no prescription for how to do this.  The issue of unitarity bounds is discussed in Section \ref{sec:unitarity}.  There is no prescription for how to use the vertex function approach in loop calculations.  Thus all the virtues of effective field theory that the Lagrangian approach lacks are also lacking in the vertex function approach.

\bigskip\bigskip

It is not uncommon to see an approach used which confuses the Lagrangian and vertex function approaches.  A Lagrangian such as Eq.~(\ref{eq:L}) is written down, but the parameters $g_1^Z, \kappa_\gamma,$ {\it etc.} are treated as form factors, that is, as functions of $P^2$.  This makes no sense, as the Lagrangian is written in position space, not momentum space, so the Lagrangian parameters cannot be functions of momentum.  This approach has all the deficiencies of the Lagrangian and vertex function approaches and should be abandoned.

\section{Anomalous Couplings from Effective Field Theory}\label{sec:anomalouseft}

The effective field theory approach allows us to reframe some analyses that have been performed using the anomalous coupling formalism.  If the anomalous couplings were taken to be constant Lagrangian parameters, then we can reinterpret them as the coefficients of dimension six operators.  By reframing the results in terms of dimension six operators, all of the desirable features of the effective field theory, listed in Section~\ref{sec:eft}, remain intact.

When anomalous couplings are derived from an effective field theory it is important to remember that they, like the underlying effective field theory, are only valid below the scale of new physics, $\Lambda$.  This is in stark contrast to the original use of anomalous couplings, which were regarded as being valid to arbitrarily high energy \cite{Gaemers:1978hg,Hagiwara:1986vm}.

The effective field theory approach described in the previous section allows one to calculate the parameters $g_1^Z$, $\kappa^\gamma$, {\it etc.}, in terms of the coefficients of the five dimension-six operators.  Calling these coefficients $c_{WWW}, c_W, c_B, c_{\tilde{W}WW}, c_{\tilde{W}}$, one finds \cite{Hagiwara:1993ck,Wudka:1994ny}
\begin{eqnarray}
g_1^Z & = & 1+c_W\frac{m_Z^2}{2\Lambda^2}\\
\kappa_\gamma & = & 1+(c_W+c_B)\frac{m_W^2}{2\Lambda^2}\\
\kappa_Z & = & 1+(c_W-c_B\tan^2\theta_W)\frac{m_W^2}{2\Lambda^2}\\
\lambda_\gamma & = & \lambda_Z = c_{WWW}\frac{3g^2m_W^2}{2\Lambda^2}\\
g_4^V &=& g_5^V=0\\
\tilde{\kappa}_\gamma & = &
c_{\tilde{W}}\frac{m_W^2}{2\Lambda^2}\\
\tilde{\kappa}_Z & = &
-c_{\tilde{W}}\tan^2\theta_W\frac{m_W^2}{2\Lambda^2}\\
\tilde{\lambda}_\gamma & = & \tilde{\lambda}_Z = c_{\tilde{W}WW}\frac{3g^2m_W^2}{2\Lambda^2}
\end{eqnarray}
Defining $\Delta g_1^Z = g_1^Z - 1$, $\Delta \kappa_{\gamma,Z} = \kappa_{\gamma,Z} - 1$, we find the relation \cite{Hagiwara:1993ck}
\begin{equation}
\Delta g_1^Z=\Delta \kappa_Z + \tan^2\theta_W \Delta \kappa_\gamma
\end{equation}
This, together with the relation $\lambda_\gamma = \lambda_Z$, reduces the five $C$ and $P$ violating parameters down to three.  For the $C$ and/or $P$ violating parameters, we find the relation
\begin{equation}
0=\tilde \kappa_Z + \tan^2\theta_W \tilde \kappa_\gamma
\end{equation}
This, together with the relations $\tilde\lambda_\gamma = \tilde\lambda_Z$ and $g_4^Z=g_5^Z=0$ reduces the six $C$ and/or $P$ violating parameters down to just two.  Thus the effective field theory approach not only has the many virtues that are lacking in the anomalous coupling approach, it is far simpler.  It provides a well motivated framework with a minimal set of parameters, which is often required due to the limited precision of the experiments.

These relations amongst the anomalous couplings are present because we have restricted our attention to dimension six operators, which are expected to be dominant.  If one includes dimension eight operators, one generally finds that these relations are no longer valid \cite{Hagiwara:1993ck}.  Nevertheless, one expects violations of these relations to be small.  This is an example of one of the desirable features of an effective field theory described in Section~\ref{sec:eft}.  The theory is general enough to describe all possible new physics, but provides guidance as to the most likely place to find it.

We can invert the above equations to find expressions for the coefficients of the dimension six operators in terms of the anomalous couplings:
\begin{eqnarray}
\frac{c_W}{\Lambda^2}& = & \frac{2}{M_Z^2}\Delta g_1^Z = \frac{2}{M_Z^2}(\tan^2\theta_W \Delta\kappa_\gamma + \Delta\kappa_Z) \\
\frac{c_B}{\Lambda^2}& = & \frac{2}{M_W^2}\Delta\kappa_\gamma - \frac{2}{M_Z^2}\Delta g_1^Z\\
& = & \frac{2}{\tan^2\theta_W M_Z^2}\Delta g_1^Z - \frac{2}{\sin^2\theta_W M_Z^2}\Delta\kappa_Z = \frac{2}{M_Z^2}(\Delta\kappa_\gamma - \Delta\kappa_Z)\\
\frac{c_{WWW}}{\Lambda^2}& = & \frac{2}{3g^2m_W^2}\lambda_\gamma = \frac{2}{3g^2m_W^2}\lambda_Z \\
\frac{c_{\tilde{W}}}{\Lambda^2}& = & \frac{2}{m_W^2}\tilde{\kappa}_\gamma = -\frac{2}{\tan^2\theta_Wm_W^2}\tilde{\kappa}_Z \\
\frac{c_{\tilde{W}WW}}{\Lambda^2}& = & \frac{2}{3g^2m_W^2}\tilde{\lambda}_\gamma = \frac{2}{3g^2m_W^2}\tilde{\lambda}_Z
\end{eqnarray}
These equations are only valid if the anomalous couplings are treated as constants, that is, independent of energy.  As we mentioned at the end of the previous section, it is inappropriate to treat them otherwise. These relations can be used to extract the coefficients of the dimension six operators from a variety of different processes.  This tests the assumption that the contributions from dimension eight operators are negligible.

The form factors of the vertex-function approach may also be calculated from the effective field theory.  We find
\begin{eqnarray}
f_1^\gamma & = & 1+c_{WWW}\frac{3g^2P^2}{4\Lambda^2}\\
f_1^Z & = & 1+c_W\frac{m_Z^2}{\Lambda^2}-c_{WWW}\frac{3g^2P^2}{4\Lambda^2}\\
f_2^\gamma & = & f_2^Z = c_{WWW}\frac{3g^2m_W^2}{2\Lambda^2}\\
f_3^\gamma & = & 2+(c_B+c_W)\frac{m_W^2}{2\Lambda^2}+c_{WWW}\frac{3g^2m_W^2}{2\Lambda^2}\\
f_3^Z & = &2+\left(c_W(1+\cos^2\theta_W)-c_B\sin^2\theta_W\right)\frac{m_Z^2}{2\Lambda^2}
+c_{WWW}\frac{3g^2m_W^2}{2\Lambda^2}\\
f_4^V&=&f_5^V=0\\
f_6^\gamma &=& +c_{\tilde{W}}\frac{m_W^2}{2\Lambda^2}
-c_{\tilde{W}WW}\frac{3g^2m_W^2}{2\Lambda^2}\\
f_6^Z &=&-c_{\tilde{W}}\tan^2\theta_W\frac{m_W^2}{2\Lambda^2}
-c_{\tilde{W}WW}\frac{3g^2m_W^2}{2\Lambda^2}\\
f_7^\gamma & = & f_7^Z = -c_{\tilde{W}WW}\frac{3g^2m_W^2}{4\Lambda^2}
\end{eqnarray}
The $C$ and $P$ conserving form factors are expressed in terms of the three independent parameters of the effective field theory.  In addition, the $P^2$ dependence of the form factors are found to be constant or linear.  The $C$ and/or $P$ violating form factors are constants, and are expressed in terms of just two parameters.
These results are just the momentum-space analogues of the results for the Lagrangian parameters above.

\section{Unitarity bounds}\label{sec:unitarity}

\begin{figure}
    \centering
    \includegraphics[width=5in]{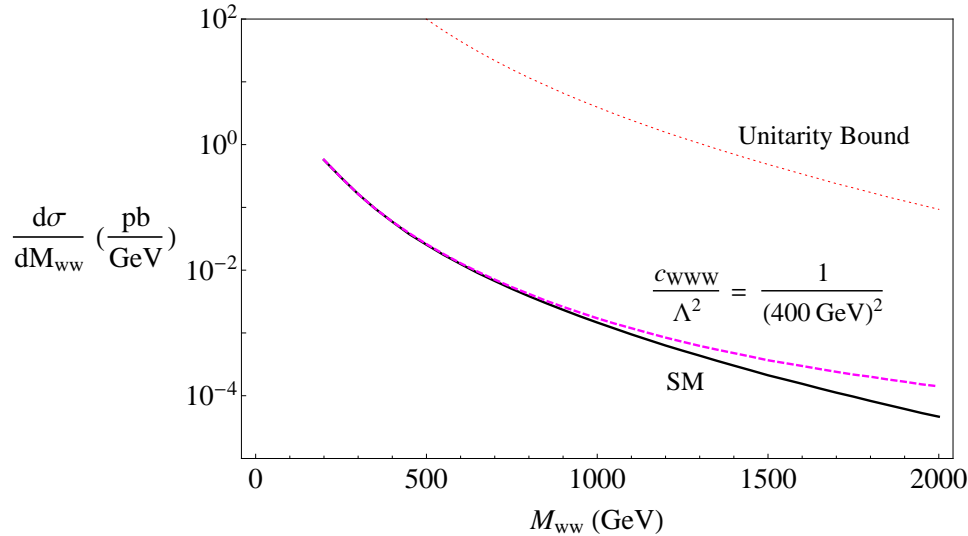}
    \caption{Differential cross section versus invariant mass for the process
    	$pp \rightarrow W^+ W^-$ at the CERN Large Hadron Collider (14 TeV).  Both the SM cross section and the cross section including the dimension six operator $O_{WWW}$ are shown.  Also shown is the unitarity bound on the cross section.
      Results were generated with Whizard \cite{Kilian:2007gr} and checked with MadGraph \cite{Alwall:2011uj}.}
    \label{lhc14_owww_400gev}
\end{figure}

Among the implications of $S$-matrix unitarity for scattering processes is an upper bound on the partial wave amplitudes.  In the case of electroweak vector boson pair production, this upper bound is far above the standard model result.  For example, we show in Fig.~\ref{lhc14_owww_400gev} the invariant mass spectrum for $W^+W^-$ production in the standard model at the CERN Large Hadron Collider (14 TeV), as well as the upper bound on the cross section obtained by saturating the $J=1$ partial wave amplitude (see Appendix~\ref{apdx:unitarity}).  We also show in Fig.~\ref{lhc14_owww_400gev} the $W^+W^-$ cross section obtained by adding the operator ${\cal O}_{WWW}$ to the standard model, with a coefficient $c_{WWW}/\Lambda^2= (400 \;{\rm GeV})^{-2}$. The cross section deviates from the standard model at large invariant mass, as expected.

In the effective-field-theory approach, the unitarity bound is irrelevant \cite{Wudka:1994ny}.  Dimension-six operators yield terms in the amplitude that grow like $s/\Lambda^2$, and will eventually violate the unitarity bound at high energy.  However, when that happens the effective field theory has become useless, and should be discarded.  Once $s/\Lambda^2$ is of order unity, there is no justification for ignoring the yet higher-dimension operators, since they are not suppressed.

In contrast, consider the anomalous couplings approach.  Since the couplings reside in the Lagrangian, they are necessarily constants (that is, independent of energy).  However, constant anomalous couplings yield amplitudes that grow like $s/M_W^2$, and eventually violate the unitarity bound at high energy.  To avoid this, one abandons the Lagrangian in favor of a vertex function, and chooses the form factors be to functions of $s$ such that they fall off at large $s$  \cite{Zeppenfeld:1987ip,Hagiwara:1989mx}.\footnote{Form factors are typically chosen proportional to $(1+s/\Lambda^2)^{-n}$, where $\Lambda$ is referred to as the form-factor or cutoff scale.  This is not to be confused with the scale $\Lambda$ of an effective field theory, Eq.~(\ref{eq:eft}).}  The unitarity bound is thus respected at arbitrarily high energies.  As we mentioned earlier, the two approaches are sometimes confused and the Lagrangian couplings are taken to be functions of $s$, which makes no sense.

This discussion highlights the conceptual difference between the effective field theory approach and that of anomalous couplings.  The effective field theory is, by construction, a low-energy theory, and has no ambitions of being applicable at arbitrarily high energy.  In contrast, anomalous couplings do not come equipped with an associated energy scale at which they should no longer be trusted.  Form factors are introduced to ensure that the unitarity bound is respected at arbitrarily high energy.  This is overly restrictive.  The only relevant constraint is that the theory respect the unitarity bound in the region where there is data.  Since the data necessarily respect the unitarity bound, an effective field theory that fits the data will automatically respect the bound.

The cross section in Fig.~\ref{lhc14_owww_400gev} is generated by adding the contribution of the dimension six operator to the amplitude, and then squaring.  It therefore is the sum of the SM cross section, the interference between the SM and the dimension-six contribution, and the square of the dimension six contribution.  For this choice of physical process and dimension six operator, the interference term is order $v^2/\Lambda^2$, and does not grow with energy.  The square of the dimension six contribution is order $s^2/\Lambda^4$, and leads to the growth of the cross section at high invariant mass.

If data were obtained that were fit by the nonstandard curve in Fig.~\ref{lhc14_owww_400gev}, one could conclude that $c_{WWW}/\Lambda^2\approx (400 \;{\rm GeV})^{-2}$.  One cannot separate the coefficient $c_{WWW}$ from the energy $\Lambda$; only the ratio $c_{WWW}/\Lambda^2$ is obtained.  However, if the data fits the curve at all measured invariant masses then $\Lambda$ must be greater than the invariant mass of the largest data point. In addition, one can conclude that $\Lambda$ must be less than the energy at which the curve violates unitarity (not shown in the figure).  Hence one could bound $\Lambda$ from above and from below.

A measurement of $c_{WWW}/\Lambda^2\approx (400 \;{\rm GeV})^{-2}$ does not indicate that $\Lambda \approx 400$ GeV.  The coefficient $c_{WWW}$ is not constrained to be unity; it can be much less or much greater.  The only constraint on $c_{WWW}$ is from the validity of perturbation theory.  The dimension six amplitude is order $c_{WWW}s/\Lambda^2$, and perturbation theory should be valid up to energies of order $\Lambda$.  At these energies the amplitude is order $c_{WWW}$; convergence of the loop expansion requires $c_{WWW} < (4\pi)^2$.

\section{Conclusions}\label{sec:conclusions}

We have argued that the time has come to abandon anomalous couplings in favor of the more modern and powerful effective field theory approach.  We discussed the many virtues of effective field theory that make it the ideal approach to nonstandard interactions of standard-model particles.  Effective field theory can be used at both tree level and loop level, and is renormalizable in the modern sense; it incorporates the gauge symmetries of the standard model; and it is general enough to describe the low-energy effects of any physics beyond the standard model, while providing guidance as to the most likely places to observe these effects.  Anomalous couplings are guaranteed to inherit these virtues only if they are derived from an effective field theory.

We applied these ideas to pair production of electroweak vector bosons.  We showed that, in addition to its many virtues, the effective field theory approach is simpler than that of anomalous couplings in that there are fewer parameters.  We also argued that unitarity bounds on cross sections are irrelevant in the effective field theory approach, providing yet another simplification.  Given its many virtues and its simplifying features, it makes good sense to adopt the effective field theory approach to nonstandard interactions of electroweak vector bosons as well as all other standard-model particles.

\section{Acknowledgements}

We are grateful for conversations with D.~Dicus and M.~Neubauer.  This work was supported in part by the U.~S.~Department of Energy under contract No.~DE-FG02-91ER40677 and by NSF grant PHY-0757889.  C.~Degrande is a fellow of the Fonds National de la Recherche Scientifique and the Belgian American Education Foundation. O.~Mattelaer is ‘Chercheur scientiﬁque logistique postdoctoral F.R.S- 
FNRS‘, Belgium.
This work was partially supported by the Belgian Federal Oﬃce for Scientiﬁc, Technical and Cultural Aﬀairs through the Interuniversity Attraction Poles Program - Belgium Science Policy P6/11-P and by the ISN MadGraph convention 4.4511.10.

\newpage

{\Large \bf Appendices}

\appendix

\section{Odd-Dimensional Operators Violate Lepton and/or Baryon Number}\label{apdx:dimodd}

\begin{table}[t]
\begin{tabular}{|c|c|c|c|c|}
\hline
																			&	Lorentz indices	&	$SU(2)$ indices	 &	Dimension	&	Total		\\\hline \vspace{-2ex}&&&&\\
$\bar{f}_Lf_R$,	$\bar{f}_Rf_L$				&	0													 &	1												 &	3					&	4				 \\\hline \vspace{-2ex}&&&&\\
$\bar{f}_L\gamma^\mu f_L$							&	1													 &	2												 &	3					&	6				 \\\hline \vspace{-2ex}&&&&\\
$\bar{f}_R\gamma^\mu f_R$							&	1													 &	0												 &	3					&	4				 \\\hline \vspace{-2ex}&&&&\\
$\bar{f}_L\sigma^{\mu\nu}f_R$,
	$\bar{f}_R\sigma^{\mu\nu}f_L$				&	2													 &	1												 &	3					&	6				 \\\hline \vspace{-2ex}&&&&\\
$\phi$, $\tilde{\phi}$								&	0													 &	1												 &	1					&	2				 \\\hline \vspace{-2ex}&&&&\\
$D^\mu$																&	1													 &	0												&	1					 &	2				 \\\hline \vspace{-2ex}&&&&\\
$B^{\mu\nu}$, $G^{\mu\nu}$, $W^{I\mu\nu}$	&	2												 &	0												 &	2					&	4				 \\\hline\hline
Effective operator				&	even											&	even										 &	$D$					&	$D$+even		\\\hline
\end{tabular}\caption{The number of Lorentz vector and $SU(2)$ fundamental indices, and the dimension, of the fields and operators.  The last row is the effective operator of dimension $D$ constructed from above fields and operators.\label{dimodd}}
\end{table}

If an effective operator conserves baryon and lepton number, the fermion fields must be paired up to form terms such as $\bar{f}_Lf_R$, $\bar{f}_L\gamma^\mu f_L$, $\bar{f}_L\sigma^{\mu\nu}f_R$, etc. There is no need to put in $\gamma_5$, because $f_L$ and $f_R$ are eigenstates of $\gamma_5$. These operators, combined with other standard model fields, are the basic ``building blocks'' of any operator.  We put these operators and fields in the first column of Table~\ref{dimodd}.

An effective operator will be composed of some combination of the operators and fields in Table~\ref{dimodd}.  Each of these operators and fields has some Lorentz indices and some $SU(2)$ fundamental representation indices, but the effective operator must be invariant under both the Lorentz and $SU(2)$ groups.  Therefore, the total number of Lorentz indices in the effective operator must be even because we need either two vectors to form a scalar or four vectors to form a pseudoscalar.  Similarly, the total number of $SU(2)$ fundamental representation indices in the operator must be even because we need two such indices to form an $SU(2)$ singlet or triplet.  These numbers are shown in the second and third columns of Table~\ref{dimodd}.  Recall that $f_L$ is an $SU(2)$ doublet and $f_R$ is a singlet.

The dimension of each of the ``building blocks'' is shown in the fourth column of Table~\ref{dimodd}.  We denote the dimension of the effective operator constructed from these ingredients by $D$.  If we add the first three numbers in each row of Table~\ref{dimodd}, the result is always an even number, so the sum of these numbers for any given lepton- and baryon-number-conserving operator must be even as well.  The sum of the last row in the Table is $D$ plus an even number.  We conclude that $D$, the dimension of the operator, must be an even number.

\section{Derivation of Cross Section Unitarity Bound}\label{apdx:unitarity}

We start with the partial wave expansion for a scattering amplitude $T$:
\begin{equation}
	T(\lambda_1 \lambda_2 \rightarrow \lambda_3 \lambda_4)
	= 16\pi \sum_{j=0}^\infty (2j+1)
	a_j D^j_{\lambda_1-\lambda_2,\lambda_3-\lambda_4}
\end{equation}
Note that the $D$-functions are orthogonal:
\begin{equation}
	\int d(\cos \theta)
	D^j_{\lambda_1-\lambda_2,\lambda_3-\lambda_4}
	D^{j^\prime \: \ast}_{\lambda_1-\lambda_2,\lambda_3-\lambda_4}
	= \frac{2}{2j+1} \delta^{jj^\prime}
\end{equation}
Thus
\begin{align}
	\int d(\cos \theta) \vert T \rvert^2 & = 2\lvert N \rvert^2 \sum_{j=0}^\infty (2j+1) \lvert a_j \rvert^2
\end{align}

Define the scattering amplitude in terms of the $S$ matrix, $S = 1 + iT$.
Unitarity requires that
\begin{align}
	\label{tunitarity}	T^\dagger T & = 2 \text{Im}(T)
\end{align}
Now take the matrix element of (\ref{tunitarity}) between identical two-body states and insert a complete set of intermediate states on the left-hand side.  This can then be written as
\begin{align}
	\int dPS_2\: \lvert T^{el} \rvert^2 + \sum_n \int dPS_n\:
	\lvert T^{in} \rvert^2 & = 2\, \text{Im}(T^{el}) \\
	\int dPS_2\: \lvert T^{el} \rvert^2 + \sum_{\lambda_3,\lambda_4}
	\int dPS_2\: \lvert T^{in} \rvert^2 & \leq 2\, \text{Im}(T^{el})
\end{align}
where we removed the sum over $n$ and selected a single 2-body inelastic scattering
process.  For massless elastic-scattering particles, the phase space integral gives
\begin{align}
	\frac{1}{16\pi} \int d(\cos \theta)\: \lvert T^{el} \rvert^2
		+ \sum_{\lambda_3,\lambda_4} \int dPS_2\: \lvert T^{in} \rvert^2
	& \leq 2\, \text{Im}(T^{el}) \\
	\sum_{j=0}^\infty (2j+1)
	\lvert a_j^{el} \rvert^2 + \frac{1}{32\pi}
	\sum_{\lambda_3,\lambda_4} \int dPS_2\: \lvert T^{in} \rvert^2
	& \leq \sum_{j=0}^\infty (2j+1) \text{Im}(a_j^{el})
\end{align}
Now we will assume that the elastic process is dominated by the $j=1$ mode and throw away all other terms:
\begin{align}
	3\lvert a_0^{el} \rvert^2 +
	\frac{1}{32\pi} \sum_{\lambda_3,\lambda_4}
	\int dPS_2\: \lvert T^{in} \rvert^2
	& \leq 3\text{Im}(a_0^{el}) \\
	\frac{1}{32\pi} \sum_{\lambda_3,\lambda_4}
	\int dPS_2\: \lvert T^{in} \rvert^2
	& \leq 3\text{Im}(a_0^{el})\left(1 - \text{Im}(a_0^{el})\right) \\
	\frac{1}{32\pi} \sum_{\lambda_3,\lambda_4}
	\int dPS_2\: \lvert T^{in} \rvert^2
	& \leq \frac{3}{4} \\
	\sum_{\lambda_3,\lambda_4}
	\int dPS_2\: \lvert T^{in} \rvert^2
	& \leq 24\pi
\end{align}
For the process $q\bar{q} \rightarrow WW$, averaging over colors and initial-state spins, we find the bound
\begin{align}
	\sigma_{tot} & = \frac{1}{2\hat{s}}
	\frac{1}{9}\sum_{colors} \frac{1}{4}\sum_{\lambda_1,\lambda_2}
	\sum_{\lambda_3,\lambda_4} \int dPS_2\: \lvert T^{in} \rvert^2 \\
	& \leq \frac{2\pi}{\hat{s}}
\end{align}


\end{document}